# tttrlib: modular software for integrating fluorescence spectroscopy, imaging, and molecular modeling

Thomas-Otavio Peulen[1,2,3,4,*,#], Katherina Hemmen[4,*], Annemarie Greife[5], Benjamin M. Webb[1,2,3], Suren Felekyan[5], Andrej Sali[1,2,3], Claus A. M. Seidel[5], Hugo Sanabria[6], Katrin G. Heinze[4#]

[1]Department of Bioengineering and Therapeutic Sciences, University of California, San Francisco, San Francisco, California, United States; [2]Department of Pharmaceutical Chemistry, University of California, San Francisco, San Francisco, California, United States; [3]Quantitative Biosciences Institute (QBI), University of California, San Francisco, San Francisco, California, United States; [4]Rudolf Virchow Center for Integrative and Translational Bioimaging, University of Würzburg, Würzburg, Germany; [5]Chair of Molecular Physical Chemistry, Heinrich-Heine University, Düsseldorf, Germany; [6]Department of Physics & Astronomy, Clemson University, Clemson, SC 29634, USA;

* First author

# Corresponding author

# Abstract

**Summary**

We introduce software for reading, writing and processing fluorescence single-molecule and image spectroscopy data and developing analysis pipelines to unify various spectroscopic analysis tools. Our software can be used for processing multiple experiment types, *e.g.*, for time-resolved single-molecule (sm) spectroscopy, laser scanning microscopy, fluorescence correlation spectroscopy, and image correlation spectroscopy. The software is file format agnostic, processes and outputs multiple time-resolved data formats. Our software eliminates the need for data conversion and mitigates data archiving issues.

**Availability and implementation**

Our software tttrlib is distributed via pip (https://pypi.org/project/tttrlib/) and as prebuilt packages for conda environments (https://anaconda.org/tpeulen/tttrlib). The underlying code is open-source and available via GitHub (https://github.com/fluorescence-tools/tttrlib). In examples presented here and in additional documentation (available at https://docs.peulen.xyz/tttrlib), we show how to implement in vitro spectroscopy and live cell image spectroscopy analysis.

# Introduction

Making raw intensity images accessible is an ongoing endeavor addressed by the tagged-image file format (TIFF), open-microscopy environment (OME) TIFF (*1*) and, more recently, by OME-Zarr (*2*). There are two challenges in making data accessible: (*i*) metadata necessary to interpret the original data must be accessible and preserved, and (*ii*) original data must be readable.

The development of the introduced software for time-resolved single-molecule and imaging data was motivated by the need for standard data interfaces, requirements for integrative modeling (*3*), and the demand for automated analysis pipelines. The lack of a vendor-independent standard time-resolved single-photon data formats was recognized by the FRET community (www.fret.community), a scientific community founded to enhance dissemination and community-driven development of analysis tools (*4*), as a major challenge for fluorescence-based integrative models, as the PDB-Dev archiving system for integrative models (*5*) recommends archiving models themselves along with all relevant experimental data and metadata and experimental and computational protocols (*6*). In single-molecule Förster resonance energy transfer experiments (smFRET) the lack of data formats was mitigated by a Hierarchical Data Format (HDF5) based format (*7*). Our software that provides an open unified interface to raw proprietary data formats led to publications with a focus on spectroscopy (*8*, *9*) and image spectroscopy (*10*) and solved challenges of an imaging facility in processing, archiving and storing spectroscopic and image spectroscopy data that needs to be openly accessible to customers.

Compared to intensity-based experiments, data types are more complex in time-resolved spectroscopy and spectroscopic imaging since spectroscopic and image information are jointly encoded (**Fig.1A**). Such information can be used in Förster resonance energy transfer (FRET) and photo-induced electron transfer (PET) experiments to map distances between fluorophores in the range from 1 to 10 nm at Ångstrom resolution (*11–15*) and inform on dynamics of molecular processes in solution and in living cells. While fluorescence correlation spectroscopy (FCS) (*16*) combined with FRET or PET can map distances as a function of time over 10 time decades with sub-nanosecond resolution (*17*), fluorescence experiments are sufficiently sensitive (*18–20*) to study single molecules for disentangling complex biomolecular systems (*21*, *22*). Fluorescence spectroscopy combined with microscopy informs on large biomolecular assembly structures either *in vitro* or in living cells (*23–28*) at any size, ranging from polyproline oligomers (*11*, *29–31*) to ribosomes (*32*) and can map interactions in living cells at high-throughput (*33*).

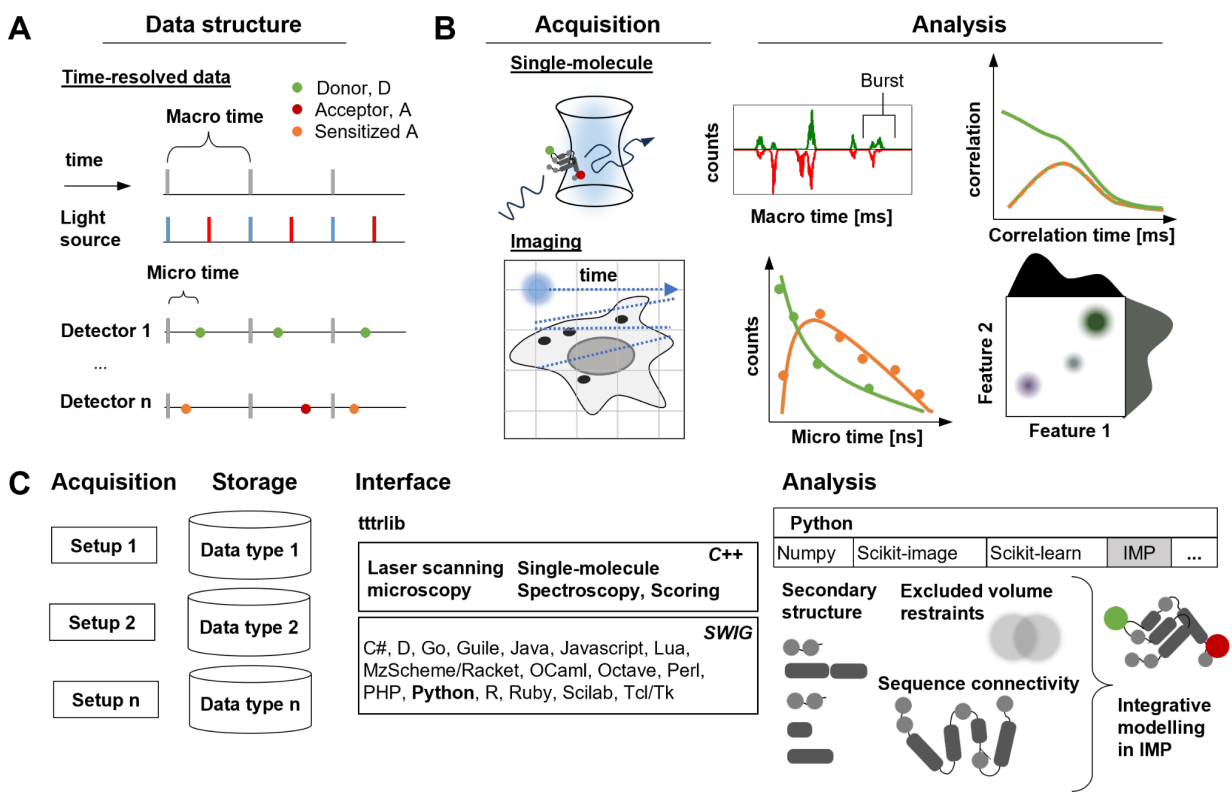

**Fig. 1** Time-resolved single-molecule fluorescence spectroscopy (SMS) and fluorescence image spectroscopy (FIS) share data registration and processing routines. (**A**) In time-resolved fluorescence, excitation and photon detection the events are encoded using a macro and a micro time counter. In pulsed experiments the coarse macro time counter is synchronized to the light source. The micro time measures the time delay relative to a macro time event (laser pulse). Additionally, events are characterized by an identifier, usually referring to the detection channel or TCSPC board input number. (**B**) *Acquisition*: Confocal SMS registers photons of freely diffusing molecules in a stationary excitation and detection volume. Confocal FIS generates an image by sweeping an excitation source over a sample and registering the emitted photons over time in a photon stream. *Analysis*: SMS and FIS process macro and micro times. In SMS, photons are grouped into single-molecule events. FIS groups photons into pixels. Macro/micro times of photons in a group are analyzed to determine spectroscopic features (*e.g.*, fluorescence lifetimes, fluorescence anisotropies). (**C**) Our C++ library tttrlib is wrapped for scripting languages via the Simplified Wrapper and Interface Generator (SWIG) abstracts data to provide an interface to data, methods, and algorithms for laser scanning microscopy and single-molecule spectroscopy. This enables fluorescence data analysis by diverse sources using other data processing and analysis libraries such as scikit-image and scikit-learn and integrative modeling by the Integrative Modeling Platform (IMP). The acquired information/data is interpreted by modeling that can consider additional information, *e.g.*, secondary structure, excluded volume or sequence connectivity information to produce a structural integrative model.

SMS and fluorescence imaging spectroscopy (FIS) rely on the time-resolved registration of photons. FIS and SMS data are encoded as photon streams (**Fig.1A**). In SMS, fluorescence bursts are selected by intensity thresholding the background. In FIS, the data stream is sorted into pixels of an image. In both cases, photons grouped into bursts or pixels need to be analyzed, *e.g.*, by counting, distribution, or correlation analysis. Frequently bursts and pixels are grouped by spatial (*e.g.*, shapes in images), temporal (*e.g.*, kymographs in time-series analysis), or spectroscopic features (*e.g.*, fluorescence lifetimes) into sub-ensembles and regions, respectively. SMS and FIS share similar challenges, as the number of photons in a pixel or burst is low (in the order of 10's to hundreds of photons). Thus, identical analysis routines are used for apparently different tasks (**Fig.1B**). Thus, even though data storage and processing routines of SMS and time-resolved are largely identical, to the best of our knowledge, no established unified framework exists for FIS and SMS data.

Therefore, we introduce a software package that preserves metadata and gives access to raw data of different proprietary data formats (*e.g.*, PicoQuant, Becker&Hickl, Zeiss, Leica) via a unified interface. Our software reads and processes spectroscopic and image data for downstream processing by software and for integrating experimental data into advanced analysis frameworks as the Integrative Modeling Platform (*34*) (IMP) (**Fig.1C**). In application examples available at https://docs.peulen.xyz/tttrlib, we illustrate how to score molecular structures for integrative modeling of molecular complexes in living cells. First, we present an SMS workflow that processes human Guanylate binding protein 1 (hGBP1) single-molecule FRET data by burst-integrated fluorescence lifetime analysis (BIFL), fluorescence correlation spectroscopy (FCS) and photon distribution analysis (PDA). We present the software's FIS capabilities, by preprocessing the data of guanylate binding protein (GBP) complexes in a workflow that combines classic image analysis with image spectroscopy.

## Implementation

*tttrlib* is a user-friendly, file format-independent interface for single-molecule and image spectroscopy data. For that, *tttrlib* mimics the expected behavior from common numeric libraries such as Numpy (https://www.numpy.org). The functionalities were implemented following standard design guidelines for operations on sequence and list data in scripting languages such as slicing of data objects. *tttrlib* objects can be initialized via keyword arguments. This allows conveniently creating and archiving settings used in analysis in non-relational object databases or dictionary files using formats such as JSON (JavaScript Object Notation), YAML (YAML Ain't Markup Language), or mmCIF (Crystallographic Information Framework), a broader system of exchange protocols based on data dictionaries and relational rules expressible in different machine-readable manifestations archived and distributed by the Protein Data Bank (PDB) and the prototype system for structural models obtained using integrative or hybrid modeling (PDB-Dev).

The time-tagged time-resolved (TTTR) data is accessed by the TTTR class. TTTR objects can be sliced and merged to facilitate or distribute the processing of larger datasets. Slicing and merging TTTR objects allows users to build filters that select events by micro time or macro time counter values to discriminate the depletion pulse in STED microscopy or molecular aggregates in single-molecule spectroscopy. Metadata of TTTR objects is accessed via a Header class and dictionaries. Modifying the metadata and saving TTTR objects to other file types enables cross-compatibility of manufacture specific analysis software. A number of selection algorithms allow defining ranges based on average intensities in a time window, and detector numbers can be used to slice and partition data. For these common operations a set of selection algorithms (intensity thresholds, selection of detection channels, etc.) and range definitions are predefined to effectively select subsets in the slicing of data. Similar applications are possible, where the user or developer has complete control.

*Requirements.* The *tttrlib* is available and tested on all major operating systems (Linux, macOS, and Windows). It can be installed in conda and scripted via Python.

*Supported microscopes and data formats.* *tttrlib* can be used for confocal laser scanning microscopy (CLSM) TTTR data and implements reading routines for the most common CLSM microscopes (such as the Leica SP5/SP8, Zeiss LSM980). So far, *tttrlib* supports reading of the proprietary data format of Picoquant (PTU/HT3), Becker&Hickl (SPC-130, SPC-630), Zeiss (confocor) as well as single-molecule SM file format and the open source PhotonHDF5 format. The most common parameters necessary for interpreting CLSM TTTR data can be user-specified. As open-source software, new data readers can be contributed and incorporated by users. Thus, *tttrlib* can be used to process arbitrary TTTR and time-resolved CLSM data.

## Materials and Methods

### smFRET data set

*Sample preparation and data acquisition*

We process data of the ligand free non-farnesylated human guanylate binding protein 1 (hGBP1) (*35*). Larger single-molecule FRET datasets on hGBP1 are openly available

(https://zenodo.org/records/6534557). Briefly, cysteines were introduced at specific positions in a cysteine-free hGBP1 variant. Here, we process data of the cysteine mutant Q344C-Q525C where we attached Alexa488 and Alexa647 via maleimide labeling chemistry (*35*).

*Burst-wise photon selection and analysis*

We select the photons emitted from the freely diffusing labeled protein as fluorescence intensity peaks ("bursts") by thresholding against a minimal number of photons in a defined time window. Here we applied a minimal threshold of 60 photons per burst. The macro- and microtime of the selected photons are saved and grouped into bursts. Based on the selected bursts, the spectroscopic properties like countrate in green or red channels, fluorescence-weighted mean fluorescence lifetime or scatter-corrected anisotropy are determined. Additionally, parameters like the average FRET efficiency can be derived from the directly determined spectroscopic parameter.

*Count-rate selective fluorescence correlation spectroscopy*

In single-molecule experiments a large amount of photons stems from uncorrelated scatter contribution, *e.g.*, the measurement buffer. This reduces the correlation amplitude. We correlate the whole photon stream and to secondly correlate only high count rate sections of the continuous data stream. Here, we specified that more than 60 photons have to be present in a 10 ms time window for the photons collected during this time window to be included in the correlation routine. The selection procedure is applied to the whole data set, i.e. photons are summed over all eight measurement channels, however, the correlation routine is selectively applied to either correlate the green (pG/sG) or the red channels only (pR/sR) or to correlate the green channels with the red channels (pG,sG/pR,sR).

*Photon Distribution Analysis (PDA)*

For photon distribution analysis, the intensity traces were sliced into time windows (TW) of 1 ms, for TWs with at least 20 photons, the green-to-red intensity ratio Sg/Sr was calculated and Sg/Sr experimental frequency histograms generated. Here, we optimize parameters of model frequency histograms of a two state or three state model. Our models additionally consider a donor-only population. In our example, we fit Sg/Sr experimental histograms by optimizing the model parameters with routines provided by scipy. *tttrlib* provides programmable PDA model outputs. First, the 1-dimensional representation of experimental two-dimensional counting data (red / green, parallel / perpendicular) can be freely specified. Thus, also anisotropy data can be optimized or sampled. Second, the forward model, i.e, the model that is used to compute experimental observables (such as Sg/Sr), is programmable. Thus, arbitrary two-dimensional counting data can be described by complex model functions.

*Burst Variance Analysis (BVA)*

We demonstrate an implementation of burst variance analysis (BVA) to estimate conformational dynamics in single-molecule Förster resonance energy transfer (smFRET) experiments. In BVA, the variance of the proximity ratio is calculated for each single-molecule burst, and dynamics are detected if this variance surpasses the shot noise limit (*36*). This standard deviation is then plotted against the average proximity ratio, with the shot noise limit determined by photon count providing a lower boundary. Events exceeding this boundary are classified as dynamic, indicating potential conformational changes.

Please note, that detection of dynamics is inherently limited by the count rate (*e.g.*, 100 kHz per molecule), which can prevent observation of fast conformational dynamics. To overcome this, we additionally estimate variance using fluorescence lifetimes (which are independent of count rate) and FRET efficiency (not shown). This approach avoids the need for sub-sampling within each burst and, therefore, remains unaffected by count rate constraints, allowing more accurate detection of rapid dynamics (*37*).

**Image spectroscopy data set**

*Sample preparation*

Measurements were performed in MEF mGBP7 deficient cells stably transduced with eGFP-mGBP7 and mCh-mGBP3. Generation, culture conditions and characterization of the cell line is described in (*26*, *38*). MEF cells were seeded in fully supplemented DMEM medium and grown until 70-80% confluence in Nunc™ LabTek™ II 8-well chambers (ThermoFisher). For live cell pulsed-interleaved excitation (PIE) MFIS-FRET measurements, the medium was changed to pre-warmed FluoroBrite™ DMEM (Gibco). Cells were kept at 37°C during the measurements.

*PIE measurement*

PIE experiments were performed on a confocal laser-scanning microscope (FV1000 Olympus, Hamburg, Germany) equipped with a single photon counting electronics with picosecond time-resolution (HydraHarp 400, PicoQuant, Berlin, Germany). eGFP was excited at 488 nm with a polarized, pulsed 20 MHz diode laser (LDH-D-C-485, Pico-Quant, Berlin, Germany) using a power of 28 nW at the objective. mCherry was excited at 565 nm with a white light laser with a 20 MHz repetition rate (NKT) using a power of 175 nW at the objective. The emitted light was collected through the same objective and separated into perpendicular and parallel polarization. A narrow range of eGFPs emission spectrum (bandpass filter: HC520/35, AHF, Tübingen, Germany) was then detected by single photon avalanche detectors (PDM50-CTC, Micro Photon Devices, Bolzano, Italy). mCherry fluorescence was detected by hybrid detectors (HPMC-100-40, Becker&Hickl, Berlin, Germany, with custom designed cooling). The mCherry detection wavelength range was set by bandpass filters (HC 609/54, AHF). To measure a single cell, we chose a 256x256 pixel ROI and collected 400 frames per image with 4 μs dwell time.

*Donor mean fluorescence lifetime and Phasor plot*

We calculate the donor mean fluorescence lifetime and perform phasor analysis (*39*) of the stacked image frames. In both cases, we only consider the green channels in the prompt time window of the PIE measurement and correct for the IRF. The donor mean fluorescence lifetime is calculated for all pixels with at least 20 photons in sum, while for the phasor plot only pixels with at least 30 photons in sum were used.

*Image segmentation*

We segmented the intensity images frame-wise into three pixel classes, nucleus, cytoplasm and vesicle-like structures (VLS), using scikit-image (*40*). For segmentation of VLS and cytoplasm, the intensity sum of the green and red channels was used. For segmentation of the cytoplasm, a median filter with three pixel radius was applied to the image frames followed by thresholding using Otsu's method (*41*). Next, small holes in the cytoplasm were filled and fragments of neighboring cells removed by selecting for a minimal area. VLS were identified from surrounding cytoplasm by Gaussian smoothing of the image frames with a single pixel radius followed by Otsu thresholding. Only identified pixel regions larger than five pixels were considered as VLS. Finally, the region of the nucleus was defined. Here, the image frames of the green channel were first smoothed using a Gaussian filter with three pixel radius, followed by thresholding using Li's method (*42*, *43*). The thresholded image was dilated, holes inside the nucleus filled and small areas outside removed. Next all left-overs outside the cytoplasm, i.e. from neighboring cells, were removed by multiplying with the generated cytoplasm mask, followed by another round of dilation to make the nucleus "smoother". Finally, we generated an average nucleus by summing over all 400 frames and thresholded this projection using the method of Otsu. Next, the three masks were combined to assign each pixel in each frame mutually exclusive to a single class, i.e. all pixels inside the cytoplasm mask which belong to the VLS or nucleus were removed from the cytoplasm mask.

*Pixel-wise analysis*

The pixel masks for nucleus, cytosol and VLS were in the next step used to select the photons present in the respective regions of interest. For the photon arrival histogram, we summed the microtimes of all photons present in the ROI up while additionally selecting for the detection channels (green / red) and the time window (prompt / delay). Additionally, we evaluated per pixel in each of those ROIs the fluorescence spectroscopy parameter such as intensity ratios, effective stoichiometry, $S_{PIE}$, or the proximity ratio, PR. To achieve this, we simply multiplied our binary masked (consisting of 0's and 1's) with the intensity, fluorescence lifetime or phasor images exported above. The resulting parameters were written into a table for easy construction of one- or two-dimensional histograms using any data analysis software. Of note, please be aware that no correction factors such as direct acceptor excitation or donor crosstalk in the acceptor channels were applied, thus the calculated parameters are only related to stoichiometry and FRET efficiency but not the correct values (*4, 44*).

Example scripts and example data that implement the described analysis pipelines are published openly in Zenodo (https://zenodo.org/records/14002224).

# Results
## Spectroscopy

Our software offers high-level and low-level processing and analysis methods that can be flexibly combined to custom analysis pipelines and new analysis methodologies. The software can process and analyze ensemble fluorescence spectroscopy, SMS, and FIS data. For SMS photon traces can be filtered intuitively (*45*). High precision FRET analysis of photon histograms that takes data shot-noise explicitly into account (*46*) is enabled by scriptable photon distribution analysis (PDA) (*47, 48*). Micro time information encoded in fluorescence data streams can be analyzed using maximum likelihood estimators for resolving model parameters such as anisotropies (*49*) and fluorescence lifetimes (*50*) at low photon counts for SMS and FIS analysis of bursts, and pixels. Model-free analysis through the phasor approach visualizes heterogeneity (*39*) for mapping metabolic states in cells by FIS(*39, 51*). Fluorescence decay models consider experimental nuisances such as pile-up (*52*) and use CPU vector extensions for fast computation of complex models. FCS enables studying dynamic systems and informs on sample heterogeneity, diffusion coefficients, binding events, or intra-molecular dynamics (*53*). We compute fluorescence correlation spectroscopy curves for the registered photon stream using efficient algorithms (*54*). The implemented spectroscopic methods and algorithms can be applied to SMS and FIS data.

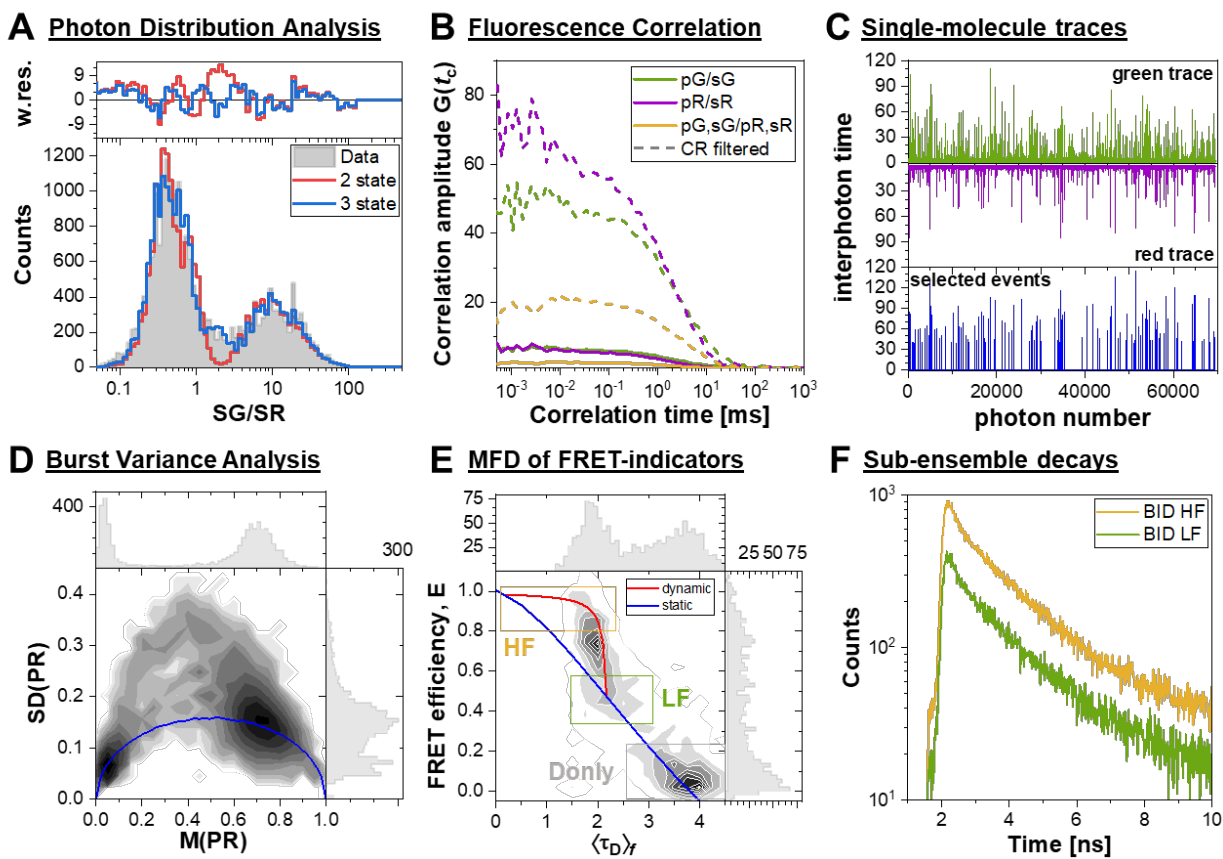

**Figure 2. Single molecule spectroscopy for dynamic structural biology on hGBP1.** **(A)** For photon distribution analysis, the intensity traces were sliced into time windows (TW) of 1 ms, for TWs with at least 20 photons, the green-to-red intensity ratio Sg/Sr was calculated and Sg/Sr frequency histograms generated. The obtained PDA histogram (gray) was fitted either with a two state (red) or three state model (blue) including an additional donor-only population. Weighted residuals are shown on top. **(B)** Fluorescence correlation curves calculated either based on the whole intensity trace (solid lines) or when selecting only high-intensity regions as shown in (B) (dashed lines). For the count-rate filtered curves, only those regions of the intensity trace were selected where more than 60 photons per 10 ms time window have been detected. Green curves show the correlation of the two donor channels (pG/sG), magenta of the two acceptor channels (pR/sR) while in yellow the cross-correlation of the donor with the acceptor channel is shown (pG,sG/pR,sR). **(C)** Short section of a single-molecule time trace of freely diffusing hGBP1 molecules labeled with Alexa488-maleimide (donor) and Alexa647-maleimide (acceptor) at positions 344 and 525 . Top two panels show the photon detection event number vs the number of photons detected in the donor (green) or acceptor (magenta) channels

per 1 μs time window. Bottom panel (blue) shows the selected photons, where a single-molecule burst was defined as the presence of more than 40 photons per 250 μs time window. **(D)** Burst variance analysis. The burst-wise mean proximity ratio, $M(PR)$, and standard deviation ($SD$) of $PR$ are shown in two-dimensional histograms (center) with one-dimensional projections of $M(PR)$ (top) and $SD(PR)$ (right). **(E)** The corrected FRET-efficiency, $E$, and fluorescence lifetime of the donor in the presence of A, $\langle \tau_D \rangle f$, are shown in two-dimensional MFD-histograms (center) with one-dimensional projections of $E$ (right) and $\langle \tau_D \rangle f$ (top). The blue and the red line are so-called FRET-line, the theoretical function connecting static, i.e., non-dynamic populations, and protein dynamics, respectively. Two double-labeled exchanging species can be identified: the low FRET species (green) and the high FRET species (yellow). **(F)** Donor fluorescence decay histograms computed for selected HF and LF bursts shown in (E).

We use a subset of the features implemented in our software in a single-molecule (sm) analysis pipeline to process human guanylate binding protein (hGBP1) smFRET data. In the confocal experiments, fluorescence of freely diffusing labeled hGBP1 in dilute solutions was registered (*35*). The analysis pipeline (*i*) reads smFRET data, (*ii*) selects single-molecule events, (*iii*) performs a burst analysis that computes intensity and lifetime-based FRET indicators, (*iv*) uses filters to correlate the photon traces, (*v*) generates single-molecule counting histograms, that are (*vi*) analyzed by photon distribution analysis (PDA), and (*vii*) selects molecular sub-ensembles,

A set of filters can be applied to the data before correlation. A photon stream can be filtered based on the macro (*55*) or the micro time information (*56–59*). We provide examples for the most common correlation approaches such as intensity filtered correlation, micro time gated correlation, and lifetime filtered correlations in the online documentation. By combining the base functionality with slicing of the photon stream into chunks with correlation methods automated robust FCS analysis can be used for live-cell measurements (*60*) and was as previously implemented for the β2-adrenergic receptor (*61*).

Using the fluorescence intensity of different detection channels fluorescence intensity counting histograms that can be used to determine FRET efficiencies (*47*) or fluorescence anisotropy (*63*) by PDA (**Fig.2A**). The photon traces obtained in a single-molecule of FCS experiment can be correlated in a software correlator that can use filters (*62*), *e.g.*, based on the photon macro and micro time (**Fig.2B**). In confocal smFRET experiments labeled molecules give rise to bursts in photon traces (**Fig.2C**). The photon traces can be binned and count rate filters are used to select windows that exceed a user-defined threshold value in the average count rate to select single molecule bursts (*45*). The photons in a burst are integrated to give average count rates. Burst-variance analysis (BVA) (*36*) can be used to identify conformational dynamics within proteins (*65*) by computing the mean and the standard deviation of (*PR*) and comparing it to the shot-noise dynamics can be identified (*37*) (**Fig.2D**). Here, we compute BVA histograms for the hGBP1 variant Q323C-Q525C. BVA highlights transitions from FRET to no FRET states (**Fig.2D**). We use the count rates in the green and the red detection channels that detect fluorescence light of donor and acceptor fluorophores in a FRET experiment to compute the FRET proximity ratio (*PR*) and the corrected FRET efficiency (*E*) for every burst (*64*). Next, we determine for every bust the fluorescence weighted average lifetimes and a corrected FRET efficiency to compute MFD histograms. The data was acquired on a MFD setup where the sample is excited by polarized light and the emitted light by the sample is split into parallel and perpendicular detection channels. Thus, we moreover analyze the fluorescence in the two channels to obtain anisotropies (*49*). In our analysis we analyze the micro times of photons in a burst using an MLE estimator to determine fluorescence averaged lifetimes (*50*). Thus, for each molecule we determine anisotropies, fluorescence lifetimes, and fluorescence intensities. Using these observables we compute multi-dimensional histograms (**Fig.3E**). Using the fluorescence observables as features, single molecules can be classified and grouped (**Fig.3E**) to resolve species by fluorescence decays of sub-ensembles (**Fig.3F**). For Q323C-Q525C we identify three populations: a high FRET population (HF), a low FRET population (LF), and molecules lacking an acceptor molecule (donor-only). The FRET molecules are not described by the static FRET line (**Fig.3E**, blue line) but is however described by a dynamic FRET line (**Fig.3E**, red line) that describes the exchange between HF and LF states (*37*). The single molecule burst can be selected based and grouped into sub-ensembles, *e.g.*, for computing sub-ensemble fluorescence decay histograms (**Fig.3F**).

**Image fluorescence spectroscopy**

We present an image spectroscopy (FIS) pipeline (**Fig.3**) that processes Multiparameter fluorescence image spectroscopy (MFIS) PIE time series acquired on MEF cells transfected with murine guanylate binding proteins 3 (mGBP3) and mGBP7 N-terminally tagged with mCherry and eGFP, respectively. GBPs are primarily localized in the cytoplasm and accumulate in vesicle-like structures (VLS) (*38, 66*). The pipeline: (*i*) groups photons into pixels of intensity images, (*ii*) performs a typical FLIM analysis for multiple detection channels and excitation sources, (*iii*) uses typical image preprocessing for segmentation of the image into pixel classes, (*iv*) uses the model free phasor approach to highlight sample heterogeneity, and (*v*) extracts fluorescence decays of pixel-classes for sub-ensemble analysis.

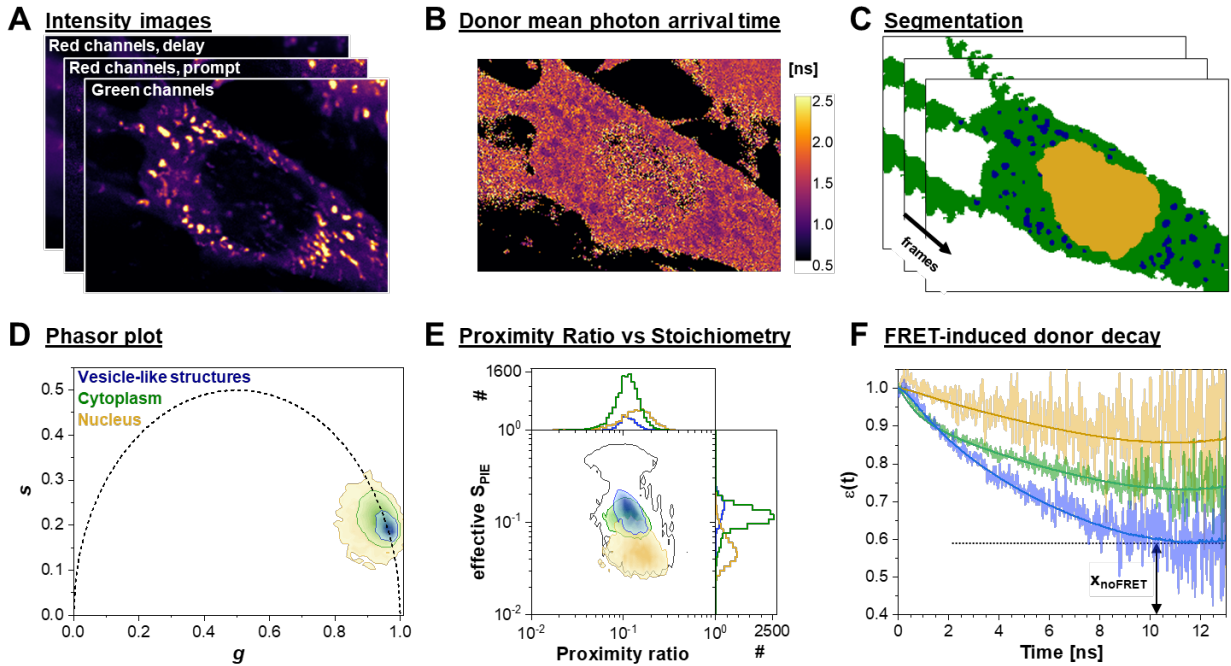

**Figure 3. PIE-MFIS FLIM dataset of an MEF cell transfected with mGBP-eGFP and mGBP-mCherry. (A)** Intensity images of the donor (eGFP; green prompt) and acceptor (mCherry; red prompt / red delay) detection channels. The total intensity is calculated from the sum of the micro time events in the specified prompt and delay time windows. Shown is the sum over all acquired frames. **(B)** Donor mean-photon arrival time calculated from the micro time distribution for the stacked frames. Only pixels with more than three detected photons per frame are considered. **(C)** Frame-wise segmentation of intensity images into three pixel classes: cytoplasm (green), vesicle-like structures (VLS, blue) and nucleus (yellow). **(D)** Phasor analysis of the donor signal (prompt time window) for the specified cellular compartments. **(E)** Pixel-wise calculation of the apparent FRET efficiency and stoichiometry for each of the three pixel classes. Marginal projections of the parameter distributions are shown on the top and right, respectively. **(F)** FRET-induced donor decay was extracted based on the photon arrival time histograms of the pixel classes defined in (E) and a reference measurement of an eGFP-only transfected cell.

We compute intensity images for the signal detected by the green detector when the sample is excited by the green light source, G|G, the red detector when the sample is excited by the green light source, R|G, and the red detector for the sample excited by the red-light source, R|R. The localization of GBPs in the cytoplasm and their accumulation in VLS is visible in intensity images (**Fig.3A**). For the photons detected in the green channel and each pixel we determine an average fluorescence lifetime, $\tau_G$, (*50*) (**Fig.3B**) and the anisotropy, $r_G$ (*49*). These features can be used for classifying pixels (*67*). We classify each pixel in each frame to either the cytoplasm (green), VLS (blue), or the nucleus (green) to study FRET location specific (**Fig.3C**). The model free phasor approach highlights sample heterogeneities (**Fig.3D**). Such heterogeneities are expected as GBPs form higher-order oligomers in VLS and are monomeric and dimeric in the cytoplasm (*26*). For each pixel in a pixel class, we compute the proximity ratio and the effective stoichiometry.

Proximity ratios and effective stoichiometries are computed using the uncorrected G|G, R|G, and R|R signal intensities. The effective stoichiometry, $S^{PIE}$, is a measure for the donor/acceptor ratio and the proximity ratio relates to the FRET efficiency (*44*). For the cytoplasm (green), VLS (blue), and the nucleus (green) we find distinct differences (**Fig.3E**). Fluorescence decays computed for entire images or for groups of pixels can be computed for fluorescence decay analysis and visualized using the extracted FRET-induced donor decay, $\varepsilon(t)$. Fluorescence decays of corresponding pixel classes can serve as input for later integrative modeling (*68*).

## Discussion

We introduced open-source software for processing various time-resolved data via a unified interface. Our software *tttrlib* integrates into various scripting languages for simple usage for custom data processing pipelines and automated image or single-molecule spectroscopy data processing. Our thin abstraction layer to vendor-specific original data and metadata addresses the challenge of preserving and processing data independent of its origin (**Fig.1C**). The customizable vendor-agnostic reading routines process data of multiple experimental setups of varying types without the need for data conversion. By operating on original data, we assert that metadata essential for handling microscopy data is preserved. Our software is programmed in C/C++ and comes with a set of algorithms and tools that operate on the ingested data. It was tested for single-molecule (**Fig.2**) and image data (**Fig.3**) registered by multiple excitation and detection modalities, including pulsed interleaved excitation (PIE) (*69*, *70*) and multiparameter fluorescence detection (MFD) (*71*) for studying protein structures and dynamics *in vitro* (*4*, *44*) and in living cells (*24*, *72*). *tttrlib* provides the most commonly used data reduction and analysis algorithms of fluorescence spectroscopy such as PDA (*47*) and correlation algorithms for FCS (*54*) that can be used for advanced FCS techniques such as full-, gated-, or filtered-FCS and methods for time-resolved image spectroscopy and image correlation spectroscopy (ICS) (*73*). Here, we presented the most common single-molecule spectroscopy and imaging modalities applications. Applications for less commonly used methods (e.g. ICS) are presented in the online repository.

We wrapped our software using the Simplified Wrapper and Interface Generator (SWIG) for simple integration into scripting and other programming languages. Currently we focus on the most widely used programming language Python. Integrating our software into Python enables the use of time-resolved data in data-science software packages such as scikit-image (*40*), scikit-learn (*74*),(*75*), and the Integrative Modeling Platform (IMP) (*34*). For easy construction of custom analysis pipelines, we implemented our software in a fully modular fashion (**Fig.1C**). This is illustrated by Python implementations of example smFRET and image spectroscopy analysis pipelines and more extensive documentation, benchmarks, and examples in the online documentation (**Implementation**). These code examples can serve as templates to develop custom analysis pipelines and data processing workflows.

Various FLIM and SMS software (lifetime fits and phasor analysis) exist. FLIMJ, based on the FLIMlib library, focuses on ease of use by providing graphical user interfaces (GUI) for basic FLIM analysis in ImageJ for Linux, macOS, and Windows (*76*). FLIMfit offers many model functions for FLIM but lacks, as stand-alone software, a close integration with image analysis (*77*). General SMS software such as PAM (*78*) and software maintained by single-molecule labs usually offer ease-of-use through GUIs and a large toolbox for FIS and SMS (*64*, *72*). This software is often based on closed-source programming languages, lacks scripting capabilities or interfaces to other toolboxes (*e.g.*, for machine learning), and is tightly integrated with GUIs (*78*). A collection of SMS software is compiled and maintained by the FRET community (see: https://fret.community/software/). More recent open-source SMS software such as PhotonHDF (*7*) and FRETBursts (*79*) mitigate the absence of open standards in SMS and give scripting capabilities and interfaces to other toolboxes; however, they require a conversion of the original data to an open format (*7*) and lack imaging capabilities. Our modular library was developed for programming analysis workflows while most existing FIS and SMS software was designed for established data processing workflows and user specific cases. By focusing on developers and data scientists, *tttrlib* can be closely integrated with other software such as NumPy/SciPy, scikit-image, or the Integrative Modeling Platform (*34*).